\documentclass[journal=jacsat,manuscript=article]{achemso}

\usepackage{chemformula} % Formula subscripts using \ch{}
\usepackage[T1]{fontenc} % Use modern font encodings
\usepackage{xcolor}

\author{Sarath Chandra Varma}
\alsoaffiliation{Department of Mechanical Engineering, Indian Institute of Science, Bangalore, India}
\author{Abhineet Singh Rajput}
\alsoaffiliation{Department of Mechanical Engineering, Indian Institute of Science, Bangalore, India}
\author{Aloke Kumar}
\alsoaffiliation{Department of Mechanical Engineering, Indian Institute of Science, Bangalore, India}
\email{*alokekumar@iisc.ac.in}
\title[An \textsf{achemso} demo]
  {Rheocoalescence: Relaxation time through coalescence of droplets}

\abbreviations{IR,NMR,UV}

\begin{document}

%%%%%%%%%%%%%%%%%%%%%%%%%%%%%%%%%%%%%%%%%%%%%%%%%%%%%%%%%%%%%%%%%%%%%
%% The "tocentry" environment can be used to create an entry for the
%% graphical table of contents. It is given here as some journals
%% require that it is printed as part of the abstract page. It will
%% be automatically moved as appropriate.
%%%%%%%%%%%%%%%%%%%%%%%%%%%%%%%%%%%%%%%%%%%%%%%%%%%%%%%%%%%%%%%%%%%%%
%\begin{tocentry}

%Some journals require a graphical entry for the Table of Contents.
%This should be laid out ``print ready'' so that the sizing of the
%text is correct.

%Inside the \texttt{tocentry} environment, the font used is Helvetica
%8\,pt, as required by \emph{Journal of the American Chemical
%Society}.

%The surrounding frame is 9\,cm by 3.5\,cm, which is the maximum
%permitted for  \emph{Journal of the American Chemical Society}
%graphical table of content entries. The box will not resize if the
%content is too big: instead it will overflow the edge of the box.

%This box and the associated title will always be printed on a
%separate page at the end of the document.

%\end{tocentry}

%%%%%%%%%%%%%%%%%%%%%%%%%%%%%%%%%%%%%%%%%%%%%%%%%%%%%%%%%%%%%%%%%%%%%
%% The abstract environment will automatically gobble the contents
%% if an abstract is not used by the target journal.
%%%%%%%%%%%%%%%%%%%%%%%%%%%%%%%%%%%%%%%%%%%%%%%%%%%%%%%%%%%%%%%%%%%%%
\begin{abstract}
Dynamics of the pendant drop coalescing with a sessile drop to form a single daughter droplet is known to form a bridge. The bridge evolution begins with a point contact between the two drops leading to a liquid neck of size comparable to the diameter of the drops. To probe this phenomenon in polymeric fluids, we quantify the neck radius growth during coalescence using high speed imaging. In the current study, we unveil the existence of three regimes on basis of concentration ratio $c/c^*$ namely, inertio-elastic $c/c^*<c_e/c^*$, viscoelastic $c_e/c^*<c/c^*<20$ and elasticity dominated regimes $c/c^*>20$. Our results suggest that the neck radius growth with time (t) obeys a power-law behaviour $t^b$, such that the coefficient $b$ has a steady value in inertio-elastic and viscoelastic regimes, with a monotonic decrease in elasticity dominated regime. Based on this dependence of $b$ on concentration ratios, we propose a new measurement technique Rheocoalescence to obtain the relaxation time of these fluids. We also show a deviation from universality proposed in literature for the elasticity dominated regime.
\end{abstract}

%%%%%%%%%%%%%%%%%%%%%%%%%%%%%%%%%%%%%%%%%%%%%%%%%%%%%%%%%%%%%%%%%%%%%
%% Start the main part of the manuscript here.
%%%%%%%%%%%%%%%%%%%%%%%%%%%%%%%%%%%%%%%%%%%%%%%%%%%%%%%%%%%%%%%%%%%%%
\section{Introduction}
Coalescence is a singular event in which two or more drops merge to form a single daughter droplet \cite{76}. The dynamics of this singular event  is governed by the liquid bridge formation and its growth. This temporal growth bears the signature of the underlying governing equation\cite{16}. Such natural processes are observed in raindrop condensation\cite{1,3} and industrial processes such as paint spray coatings\cite{5,6}, combustion process\cite{14}, droplets on surfaces\cite{49}, and processes linked to life \cite{7,10}. Depending on the relative orientation of droplets, the phenomenon can occur in physically different configurations, i.e., pendant-pendant\cite{18,19,17}, sessile-pendant\cite{our}, and sessile-sessile\cite{varma2021coalescence,PRL2006,lamgmuir2012}. The entire evolution process in pendant-pendant and sessile-pendant configurations is driven by a balance between surface tension, viscous and inertial effects, and Laplace pressure\cite{16,17}. In Newtonian fluids, based on the force balance the evolution lies either in the inertial dominated\cite{21} or viscous dominated regime\cite{18,19}. Apart from these regimes, a new regime of inertially limited viscous regime\cite{67} was proposed in Newtonian droplet coalescence, wherein all inertial, viscous, and surface tension forces are essential. 

The kinematics of the coalescence phenomenon in pendant-pendant and sessile-pendant configurations is characterized by the temporal evolution of the liquid bridge of neck radius $R$ and bridge semi-width $H$. In Newtonian droplets, the temporal evolution of neck\cite{19} was demonstrated to follow the scale of $R\sim t^b$, where, $R$ is the neck radius and $t$ is time. Based on viscosity of the fluid, the dynamics of the neck radius evolution has been identified to have dominant viscous regime at early times and inertial regime at later instances. In the viscous regime\cite{19}, the neck radius has a scaling of $R\sim t$. Similarly in the inertial regime\cite{19,21}, neck radius has a scale of $R\sim t^{0.5}$. In literature, regime-wise universality \cite{16,17,18,19,20,21,22,23,24,25,26,27,28, 29,30,31,32,44} is elucidated both experimentally and analytically. In the viscous regime\cite{21}, the neck radius has a universal scaling of $R^*\sim(t^*)$, in which $R_c=R_o$ and $t_c=\eta R_o/\sigma$, where $R_o$ is radius of the drop, $\eta$ is viscosity and $\sigma$ is surface tension. Similarly, in the inertial regime\cite{16,21} neck radius has a universal scale of $R^*\sim(t^*)^{0.5}$, in which $R_c=R_o$ and $t_c=\sqrt{\rho {R_o}^3/\sigma}$, $\rho$ being density. 

The paradigm of a coalescence phenomenon in rheologically complex fluids is significantly more involved. Polymeric fluids are a distinct subgroup of complex fluids that exhibit strong non-Newtonian characteristics due to molecular chain interactions or relaxations. Relaxation time ($\lambda$) is the fingerprint of elasticity and molecular relaxations. A recent study on aqueous solutions of polymer droplets on both pendant-sessile\cite{our} and sessile-sessile\cite{varma2021coalescence} configurations emphasized the role of relaxation time on the dynamics of neck radius evolution. The former study on pendant-sessile\cite{our} configuration showed that for $Wi\sim\mathcal{O}(1)$, where $Wi=\lambda U/R$ ($\lambda$ is relaxation time, $U$ is neck velocity) is Weissenberg number, the neck radius growth follows the scale of $R\sim t^{0.36}$. The study also showed that for $Wi\sim\mathcal{O}(10^{-3}-10^{-4})$, the neck radius growth follows the scale of $R\sim t^{0.39}$. The study also showed the universality in the coalescence of polymeric droplets by non-dimensionalising the neck radius and time with $R_c=\sqrt{\nu_o\lambda}$ and $t_c=\mathrm{Oh}\lambda(c/c^*)^{1.2}$, respectively, where, $\nu_o$ is the kinematic viscosity of the fluid, $\lambda$ is relaxation time, $\mathrm{Oh}$ is Ohnesorge number and $c/c^*$ is concentration ratio, has the universal scaling of $R^*\sim {t^*}^{0.36}$. Similar to the relaxation time, concentration ratio $c/c^*$ is another important parameter representing the chain entanglements. Previous studies on coalescence of polymeric droplets were done on the solutions of $c/c^*<10$ \cite{our}. In the present study we investigate the coalescence of the polymeric droplets with $c/c^*>10$.

Despite of many applications of coalescence of polymeric droplets in microfluidics and interficial rheology\cite{48,55}, this phenomenon is sparsely studied. In the present study, we demonstrate that the coalescence of sessile and hanging pendant drops of aqueous polymer solutions have different regimes, along with the dependence of neck growth on relaxation time. To experimentally depict the effect of relaxation time on neck growth, we study the coalescence of droplets for various concentrations of polyethylene oxide (PEO) of molecular weights $M_w=5\times10^6$ g/mol and $M_w=4\times10^6$ g/mol. Experimental observation of neck radius growth of various concentrations is demonstrated by scaling analysis based on linear Phan-Thein-Tanner (PTT)\cite{ptt1,ptt2} constitutive equation. Our results contrast the universal behaviour proposed previously and hold enormous promise for opening a new method to determine the relaxation time of the fluid. 

\section{Materials and methods}
Polyethylene oxide (PEO) of different molecular weights $M_w$ are added to DI water in sufficient quantities to get the various concentrations $c$. All the solutions are stirred at 300 RPM for different durations. Polymers used in the present study along with their molecular weights are listed in Table-I. Concentrations of the polymers are chosen in a way that the solution types vary in a range of semi-dilute unentangled, and semi-dilute entangled regimes. Regimes of semi-dilute unentangled, and semi-dilute entangled are differentiated using the critical concentration $c^*$ and the entanglement concentration $c_e$ respectively. The critical concentration of PEO for different molecular weights is obtained from the $[\eta]$ intrinsic viscosity using the Flory relation $\displaystyle c^*=1/[\eta]$ alongside the Mark-Houwink-Sakurada correlation\cite{43} $[\eta]=0.072M_w^{0.65}$ and the entanglement concentration $c_e$ is obtained using the relation $\displaystyle c_e\approx 6c^*$\cite{63}. The values of $c^*$ and $c_e$ are listed in Table-I. All the concentrations used in present study along with their concentration ratios $c/c^*$ are given in Table-II.

\begin{table}
  \caption{List of molecular weights of polymers along with their critical and entanglement concentrations.}
  \label{tbl:example}
  \begin{tabular}{llll}
    \hline
    Polymer  & $M_w$ (g/mol)  & $c^*$ (\% w/v)  & $c_e$ (\% w/v) \\
    \hline
    PEO   & $5\times10^6$   & 0.061   & 0.366   \\
    PEO   & $4\times10^6$   & 0.071   & 0.426   \\
    \hline
  \end{tabular}
\end{table}

Experiments are performed on a Polydimethylsiloxane (PDMS) coated glass substrate. Before the experiments the substrate are cleaned with detergent followed by sonication with acetone and DI Water respectively for 20 mins each. The substrates are then dried in a hot air oven 95$^\circ$C for 30 mins. PDMS is prepared by adding the curing agent (Syl Gard 184 Silicone Elastomer Kit, Dow Corning) to PDMS in the ratio of 1:10. This mixture is agitated and kept for desiccation for 30 minutes until all visible gas bubbles are removed. Glass substrates are coated with PDMS using a spin coater at 5000 rpm for 60 s. The coated substrate are cured by keeping them in a hot air oven at 90$^\circ$C for at least 90 min.
Surface tension $\sigma$ of the solutions are measured by pendant drop method using optical contact angle measuring and contour analysis system (OCA25) instrument from Dataphysics. All the solutions were found to have surface tension values of $0.062\pm0.02$ N/m. We have assumed the density of all the solutions to be 1000 kg/m$^3$.

\begin{table*}
  \caption{Rheological properties of the solutions. (Note: Relaxation time values given in blue are obtained from the correlations, remaining values are obtained from the crossover of $G'$ and $G''$.}
  \label{tbl:example}
  \begin{tabular}{lllll}
    \hline
    $M_w$ (g/mol)  & $c$ (\% w/v)  & $c/c^*$ (\% w/v)  & $\eta_o$ (Pa.s)  & $\lambda$ (s) \\
    \hline
    & 0.1 & 1.64 & 0.006 & \color{blue}{0.0017}\\
    & 0.2 & 3.28 & 0.018 & \color{blue}{0.0025}\\
    & 0.4 & 6.56 & 0.06 &  \color{blue}{0.064}\\
    & 0.75 & 12.29 & 0.8 & \color{blue}{0.165}\\
    & 1 & 16.39 & 4.5 & 0.5\\
    & 1.5 & 24.59 & 20 & 0.67\\
    $5\times10^6$ & 1.75 & 28.68 & 40 & 1.325\\
    & 2.25 & 36.88 & 72 & 1.43\\
    & 2.5 & 40.98 & 85& 1.59\\
    & 2.75 & 45.08 & 190 & 2.0\\
    & 3 & 49.18 & 210 & 2.25\\
    & 3.25 & 53.28 & 230 & 2.5\\ 
    \hline
    & 0.5 & 8.19 & 0.038 & \textcolor{blue}{0.058}\\
    & 1.0 & 16.39 & 0.6 & 0.25\\
    $4\times10^6$ & 1.5 & 24.59 & 2 & 0.57\\
    & 2.0 & 32.79 & 10 & 1.04\\
    & 2.5 & 40.98 & 16 & 1.65\\
    & 3.0 & 49.18 & 75 & 2.4\\
    \hline
  \end{tabular}
\end{table*}

\section{Rheology}
\subsection{Rheometry}
\begin{figure}[ht]
\includegraphics[width=1\textwidth]{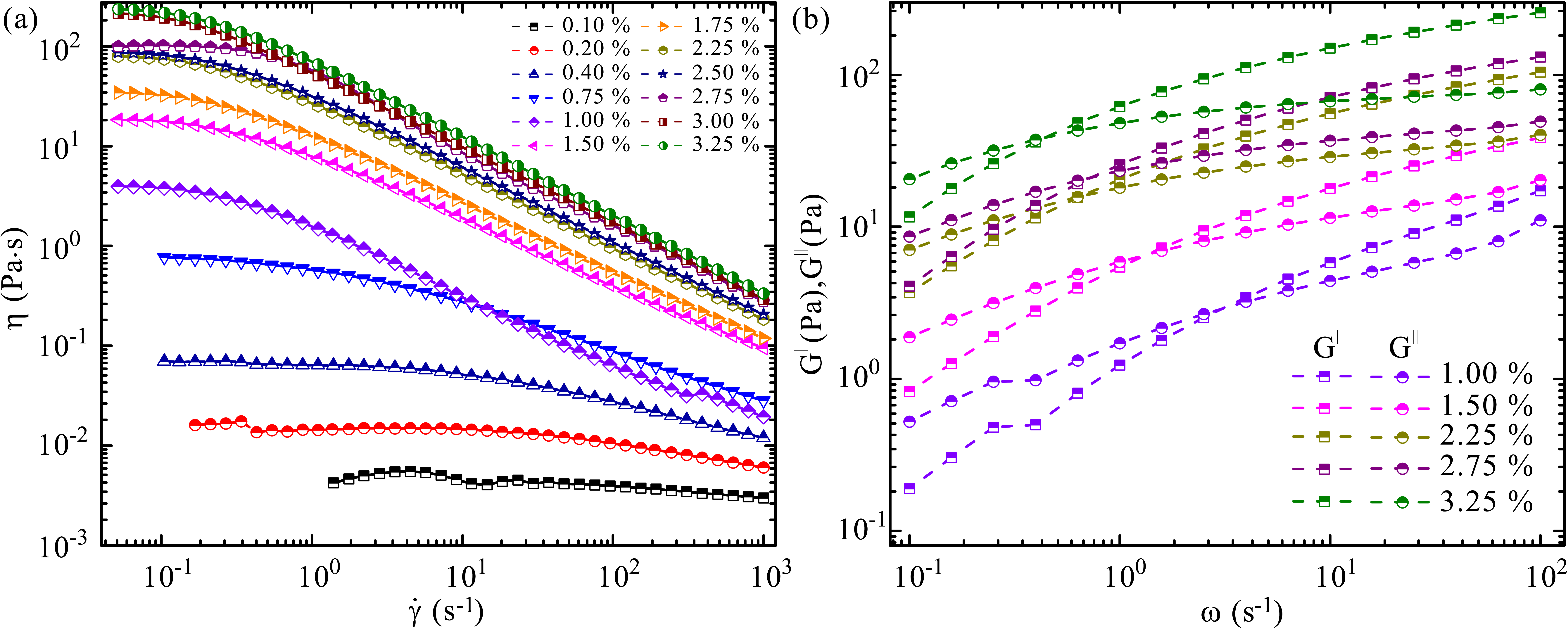} 
\caption{Rheological behavior of PEO $M_w=5\times10^6$ g/mol (a) Dependence of viscosity on shear rate for different concentrations. (b) Variation of the storage modulus $G'$ and the loss modulus $G''$ with frequency obtained from SAOS experiments for 1\% w/v, 1.5\% w/v, 2.25\% w/v, 2.75\% w/v and 3.25\% w/v concentrations. (Standard deviation of the data is less than 2\% for all the concentrations)}.
\label{fig:rheology} 
\end{figure}

Rheology experiments are performed on Anton Paar$^{\tiny{\text{\textregistered}}}$ MCR 302 rheometer using a cone and plate 40 mm, 1$^{\circ}$ geometry to characterize the viscoelastic behaviour of the solutions. The viscosity variation with shear rate for the chosen solutions is shown in Fig.~\ref{fig:rheology}(a). All the concentrations have shown the shear thinning behavior. The zero shear viscosity of the solutions is obtained by fitting the viscosity data in the form of Carreau-Yasuda model \cite{bird} represented by equation $\displaystyle \eta-\eta_{\infty}=\left(\eta_o-\eta_{\infty}\right)\left[1+\left(\Gamma \dot{\gamma}\right)^p\right]^{\frac{n-1}{p}}$, where $\displaystyle \eta_o,~\eta_{\infty},~\dot{\gamma},~n,~\Gamma$ and $\displaystyle p$ represent zero-shear viscosity, infinite-shear viscosity, shear rate, flow behavior index, time constant and width of the transition region between $\eta_o$ and the power-law region respectively. The values of $\eta_o$ for all the concentrations are listed in Table-II. Viscoelasticity of the polymer solutions is characterized by performing the small amplitude oscillatory shear SAOS experiments in rheometer. The variation of storage modulus $G'$ and loss modulus $G''$ with frequency $\displaystyle \omega$ is shown in Fig.~\ref{fig:rheology}(b) for 1\% w/v, 1.5\% w/v, 2.25\% w/v, 2.75\% w/v   and 3.25\% w/v concentrations as a representation. 

\subsection{Relaxation time}

In SAOS, relaxation time $\lambda$ of the polymer solutions is defined as $\lambda=1/\omega_c$, where $\omega_c$ is the crossover frequency for the $G'$ and $G''$ curves. It is observed that for concentrations $c>1\%$ w/v, SAOS has a crossover. But, for $c<1\%$ w/v, there is no crossover as rheometer has the maximum frequency of 100 $\textrm{s}^{-1}$ which corresponds to time scale of $10^{-2}$ s. So, for $c<1\%$ w/v, the relaxation times are estimated using the Zimm model \cite{bird}.

\begin{equation}
  \lambda_z=\frac{1}{\zeta(3\nu)}\frac{[\eta]M_w\eta_s}{\mathrm{N_A k_B} T}   
\end{equation}

where, $\eta_s$ is the solvent viscosity,  $\mathrm{k_B}$ is the Boltzmann constant, $\displaystyle \lambda_z$ is the Zimm relaxation time,  $T$ is the absolute temperature and $\nu$ is fractal polymer dimension determined using the relation $a=3\nu -1$, where $a$ is the exponent of Mark-Houwink-Sakurada correlation. For the solutions in in semi-dilute unentangled $\displaystyle \lambda_{\mathrm{SUE}}$ and semi-dilute entangled $\displaystyle \lambda_{\mathrm{SE}}$ regimes, the relaxation times are calculated using these correlations :
$\displaystyle \lambda_{\mathrm{SUE}}=\lambda_z\Big(\frac{c}{c^*}\Big)^{\frac{2-3\nu}{3\nu-1}}$ and $ \displaystyle \lambda_{\mathrm{SE}}=\lambda_z\Big(\frac{c}{c^*}\Big)^{\frac{3-3\nu}{3\nu-1}}$ 
\cite{64,65,66} respectively. The relaxation times for the chosen concentrations are listed in Table-II. The relaxation times obtained for $c>1\%$ w/v from the crossover frequency of $G'$ and $G''$ are in good agreement with Zimm model estimated values. As a representation the relaxation time obtained from the frequency sweep for 1.5\% w/v is 0.67 s, compared with the value obtained from the Zimm model as 1 s.

\section{Experiments}

\begin{figure*}[!htb]
\includegraphics[width=1\textwidth]{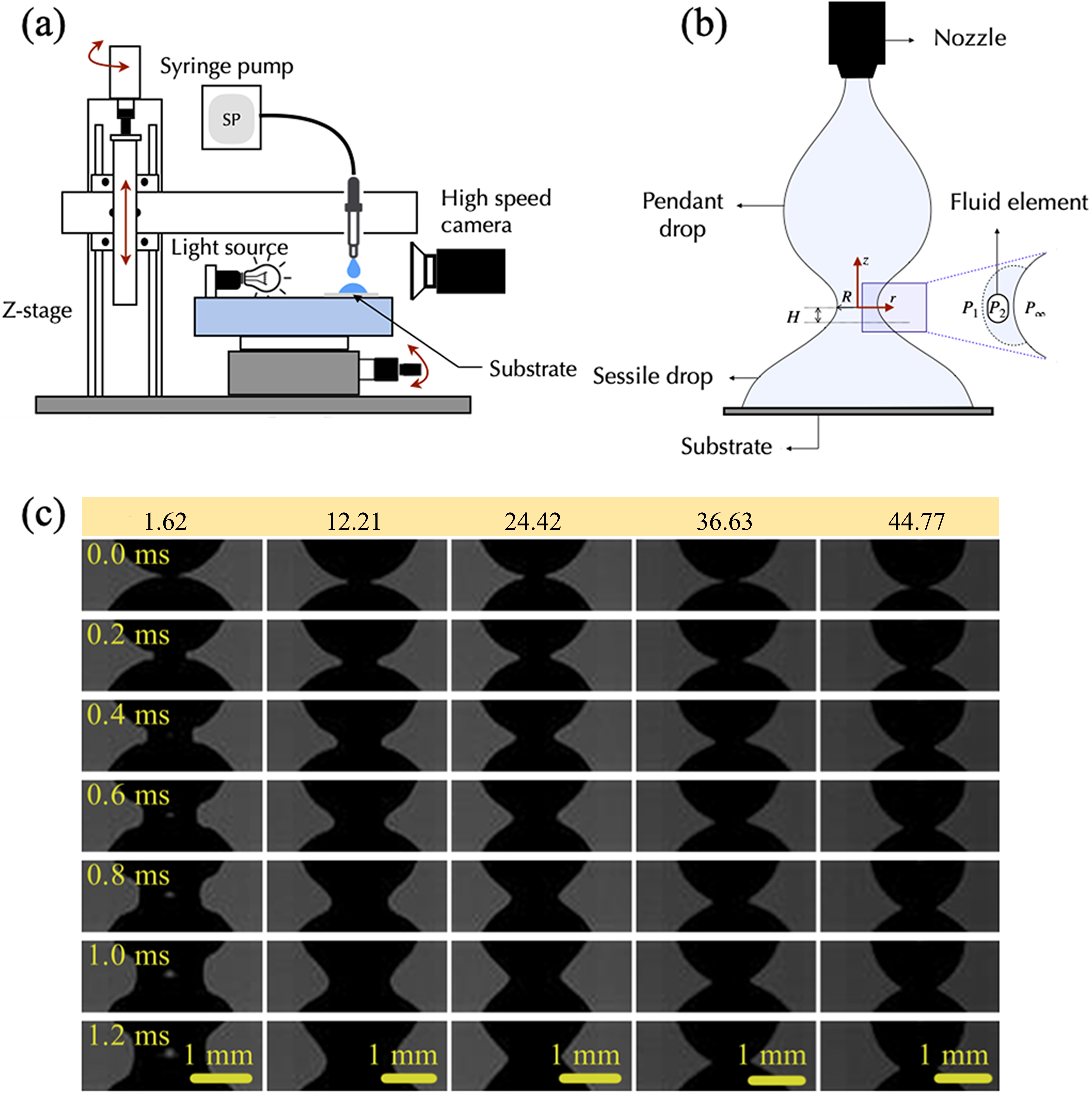} 
\caption{Schematics of (a) Experimental setup, (b) Neck region during coalescence representing the geometrical parameters during the process, (c) neck radius evolution of various concentration ratios: 1.63, 12.21, 24.42, 36.63, and 44.77 of PEO at different instants.}
\label{fig:setup} 
\end{figure*}

A drop of diameter $2.25\pm0.1$ mm is dispensed on a substrate. To achieve coalescence a pendant drop of the same diameter is brought towards the dispensed drop with $10^{-4}$ approach velocity to ensure the controlled coalescence. Experiments are conducted at a temperature of 25$^\circ$C and 1 atm pressure. Fig.~\ref{fig:setup}(a) shows the schematic of the experimental setup. The coalescence process is captured at 170000 fps using a Photron Fastcam mini high-speed camera with a Navitar lens attachment. The drops are illuminated using an LED light source. Data extraction from the images is performed using custom-written algorithms in MATLAB.

\section{Results and Discussion}
Coalescence proceeds via the formation of a liquid bridge during the merge of a pendant and sessile drop. This phenomenon is characterized by two geometric parameters namely the neck radius $R$ and the neck semi-width $H$ as shown in Fig.~\ref{fig:setup}(b). The neck radius grows with time due to the local curvature effects caused by surface tension $\sigma$. Such growth of neck radius for the concentration ratios: 1.63, 12.21, 24.42, 36.63, and 44.77 of PEO $M_w=5\times10^6$ g/mol at different time instants are shown in Fig.~\ref{fig:setup}(c). It is evident from Fig.~\ref{fig:setup}(c) that for a particular time instant, the bridge curvature for different concentration ratios has a significant change as the ratio increases.

%Coalescence is initiated immediately after the drops come into contact due to capillarity forming a liquid bridge. The geometry of liquid bridge having the neck radius $R$ and semi-bridge height $H$ is represented in Fig.~\ref{fig:setup}(b). Temporal evolution of neck radius is proceeded by the local curvature effects and surface tension $\gamma$. Such evolution of the neck radius during coalescence of an aqueous solution of 0.1\% w/v, 0.75\% w/v, 1.5\% w/v, 2.25\% w/v, and 2.75\% w/v of PEO at different time instants is shown in Fig.~\ref{fig:setup}(c). It is evident from the Fig.~\ref{fig:setup}(c) the bridge curvature has significant change in curvature as the concentration increases at a particular time instant.

\begin{figure*}[htp]
\includegraphics[scale=0.39]{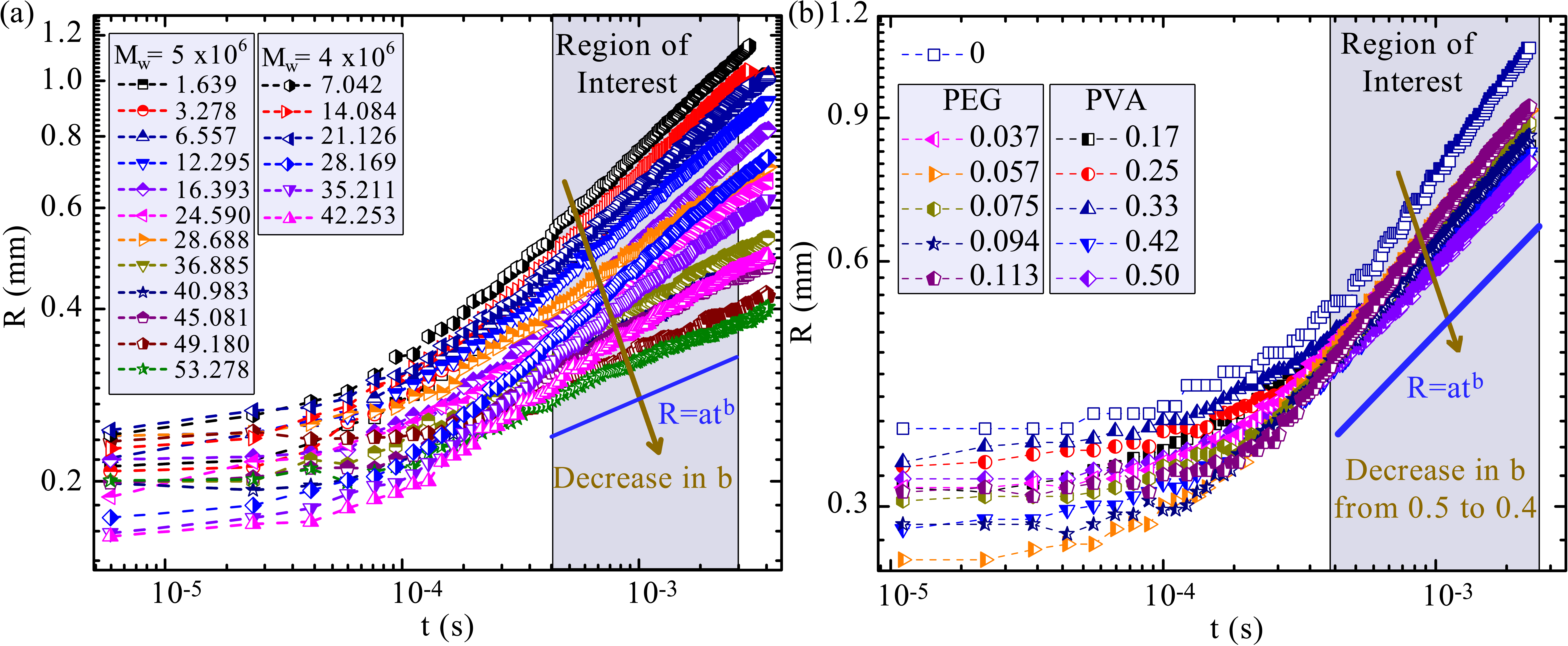} 
\caption{(a) Evolution of neck radius for various concentration ratios $c/c^*>1$ of PEO solutions showing the decrease in intercept $a$ and slope $b$ with concentration ratios. (b) Neck radius evolution for $c/c^*<1$ of Polyethylene glycol (PEG) and Polyvinyl alchol (PVA) obtained from Sarath et al.\cite{our} along with DI Water ($c/c^*=0$) representing the decrease in $b$ from 0.5 to 0.4 with addition of polymer. (Note:The error in the measurements is less than 5\%)}
\label{fig:hvst} 
\end{figure*}

The temporal evolution of the neck radius, for various concentration ratios of the polymeric drops is shown in Fig.~\ref{fig:hvst}. The neck radius growth for the the concentration ratios represented in Fig.~\ref{fig:hvst}(a) are the averaged values of 5 trials. It can be seen that the bridge has slow growth initially followed by faster growth. As previously reported in the literature, it is seen that the neck growth follows the universal power-law growth function\cite{our}, $R=at^b$ which is equivalently the linear regime in Fig.~\ref{fig:hvst}. For different concentration ratios of polymeric droplets there is a decrease in neck speed due to the change in neck curvature. This is reflected in the power law index $b$. The variation of $b$ for different concentration ratios of polymeric droplets is illustrated in Fig.~\ref{fig:hvst}(a). For $M_w=5\times10^6$ g/mol, the value of $b$ ranges from 0.38 to 0.16 while, for $M_w=4\times10^6$ g/mol the value of $b$ ranges from 0.39 to 0.25 for the range of concentration ratios explored in the current study. Fig.~\ref{fig:hvst}(b) shows the neck radius evolution of Polyethylene glycol (PEG) and Polyvinyl alcohol (PVA) obtained from Sarath et al.\cite{our} for $c/c^*<1$ along with DI Water $c/c^*=0$. It also shows the decrease in $b$ from 0.5 to 0.4 with slight addition of Polymer in DI Water. 

\begin{figure*}[h!]
\includegraphics[scale=0.53]{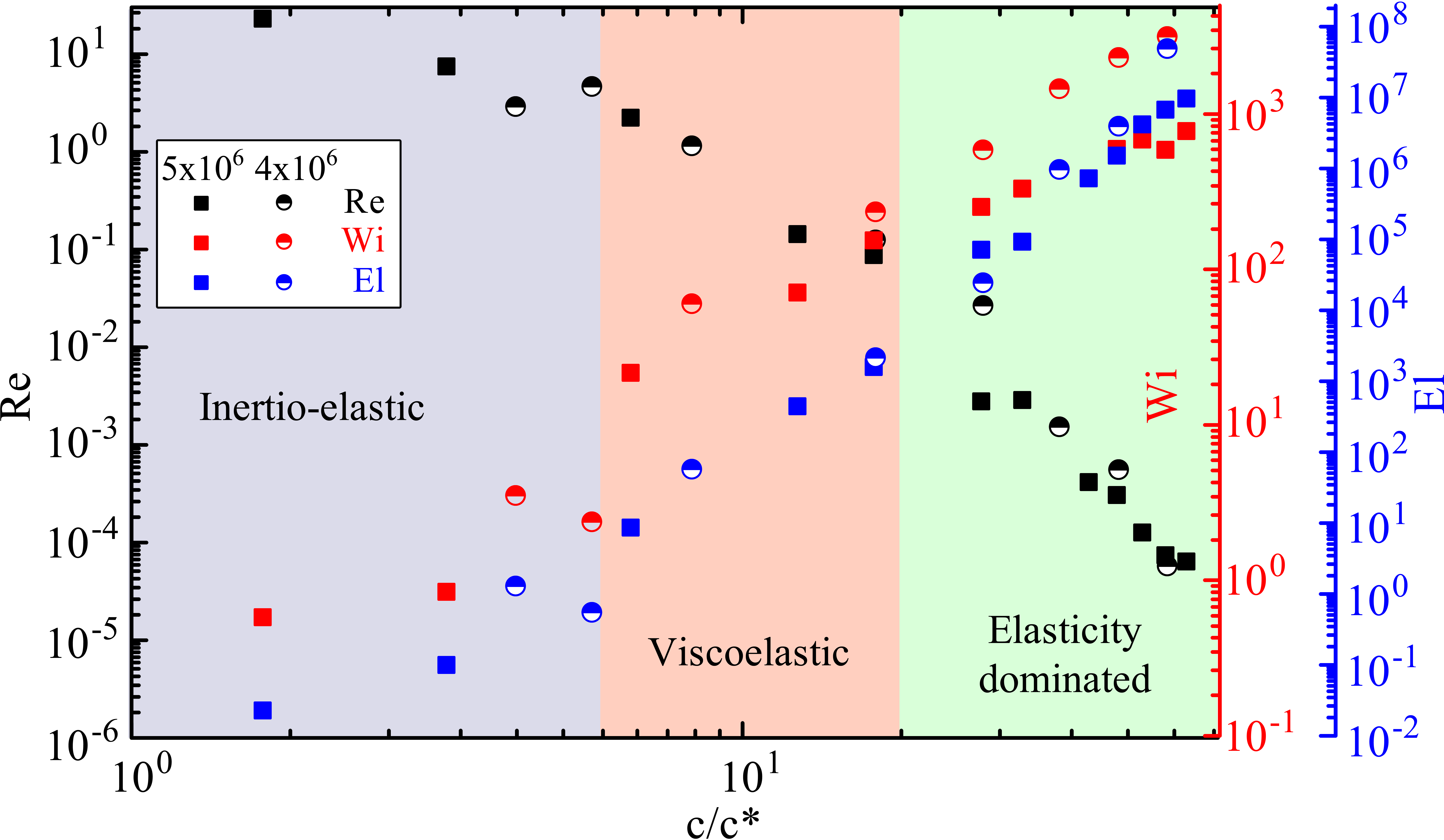} 
\caption{Comparison of predominant forces using Reynolds number, Weissenberg number, and Elastic forces represnting the inertio-elastic, viscoelastic and elasticity dominated regimes with $c/c^*$ for PEO of different molecular weights.}
\label{fig:numbers} 
\end{figure*}

To encapsulate the coalescence dynamics in polymeric fluid droplets, it is crucial to outline the underlying forces. These underlying forces are, capillary force $F_c$, inertial force $F_i$, viscous force $F_v$, and elastic force $F_e$. Among these forces, $F_c$ drives the bridge growth while the other three forces oppose it. The effect of these opposing forces $F_i$, $F_v$ and $F_e$ can be captured by three non-dimensional numbers: Reynolds number $Re=<\rho u_cl_c/\eta_o>$, $Wi=<\lambda u_c/l_c>$, and Elasticity number $El=<{\eta_o\lambda}/{\rho{l_c}^2}>$, where $u_c$ and $l_c$ represent characteristic velocity and length scales respectively, $\rho$ is density, and $\eta_o$ is zero shear viscosity. The characteristic scales associated with the flow are $u_c\sim\partial R/\partial t$ and $l_c\sim R$. The variation of these non-dimensional numbers with concentration ratio $c/c^*$ is presented in Fig.~\ref{fig:numbers}. It reveals the presence of 3 regimes based on the concentration ratios. In the first regime, with concentration ratios  $c/c^*<c_e/c^*$, the orders of corresponding numbers are $Re\sim\mathcal{O}(10)$,  $Wi\sim\mathcal{O}(10^0)$ and $El\sim\mathcal{O}(10^{-1})$ suggesting the dominance of inertia force over viscous and elastic forces i.e. $F_i>F_v\approx F_e$. As the inertial forces are predominant, this regime is an inertio-elastic coalescence. While for the second regime, with the concentration ratios $c_e/c^*<c/c^*<c_c/c^*$ ($c_c/c^*$ $\approx 20$), $Re\sim\mathcal{O}(10^{-1})$, $Wi\sim\mathcal{O}(10)$ and $El\sim\mathcal{O}(10^2)$ indicating that $F_e>F_v>F_i$. As the elastic forces are predominant followed by the viscous this regime is a viscoelastic coalescence. Similarly, for the regime with $c/c^*>c_c/c^*$, $Re<\mathcal{O}(10^{-1})$, $Wi>\mathcal{O}(10^2)$ and $El>\mathcal{O}(10^3)$ indicating that $F_e>>F_v>>F_i$. In this regime, the elastic forces are much greater than viscous forces, making it an elasticity dominant coalescence. Conclusively, as we increase the value of $c/c^*$, the coalescence phenomenon shifts from inertio-elastic to elasticity dominated regime.

The effect of the predominant forces in above 3 regimes are expounded by non-dimensionalizing, the radial $r$ direction  momentum equation under the quasi-radial assumption. The non-dimensional variables are defined as: $\upsilon^*_r=\upsilon_r/u_c$, $r^*=r/R$, $z^*=z/R$, $t^*=t/T$, $\tau^*_{rr}=\tau_{rr}/\tau_{RC}$,  $\tau^*_{rz}=\tau_{rz}/\tau_{ZC}$, $p^*=p/P_c$, where $T:=R/u_c$ and, $P_{c}:=\sigma/R_o$ ($R_o$ is the droplet radius) are the characteristic time and pressure respectively.  

%This physics of polymer chain relaxation is expounded in non-dimensionalizing the $r$-direction momentum equation under quasi-radial assumption. The non-dimensional variables are defined as: $\upsilon^*_r=\upsilon_r/u_c$, $r^*=r/R$, $t^*=t/T$, $\tau^*_{rr}=\tau_{rr}/\tau_{RC}$,  $\tau^*_{rz}=\tau_{rz}/\tau_{ZC}$, where $T:=R/u_c$ and, $\tau_{RC}:=\eta/\lambda$, $\tau_{ZC}:=\eta/\lambda$ and, $P_{c}:=\sigma/R_o$ are the characteristic stresses and pressure respectively.  

\begin{eqnarray}
\frac{\rho u_c^2}{R}\big( \frac{\partial \upsilon_{r}^*}{\partial t^*}+\upsilon_r^*\frac{\partial \upsilon_r^*}{\partial r^*} \big)=-\frac{P_c}{R}\frac{\partial p^*}{\partial r^*}+\frac{\tau_{RC}}{R}\big(\frac{\tau_{rr}^*}{r^*}+\frac{\partial\tau_{rr}^*}{\partial r^*}\big)+\frac{\tau_{ZC}}{R}\frac{\partial\tau_{rz}^*}{\partial z^*} \end{eqnarray}

The characteristic scales of stresses $\tau_{RC}$ and $\tau_{ZC}$ are obtained by introducing the previously defined non dimensional variables, along with the quasi-radial assumption in linear Phan Thein Tanner constitutive equation as follow:
\begin{eqnarray}
\frac{\partial \tau_{rr}^*}{\partial t^*}+\upsilon_r^*\frac{\partial \tau_{rr}^*}{\partial r^*}-2\tau_{rr}^*\frac{\partial \upsilon_r^*}{\partial r^*}+\frac{\tau_{rr}^*}{\frac{\lambda U}{R}}\left[1+\frac{\kappa\lambda}{\eta}\tau_{RC}\tau_{rr}^*\right]=2 \frac{\eta}{\lambda \tau_{RC}}\frac{\partial \upsilon_r^*}{\partial r^*} 
\end{eqnarray}

\begin{eqnarray}
\frac{\partial \tau_{rz}^*}{\partial t^*}+\upsilon_r^*\frac{\partial \tau_{rz}^*}{\partial r^*}-\tau_{rz}^*\frac{\partial \upsilon_r^*}{\partial r^*}+\frac{\tau_{rz}^*}{\frac{\lambda U}{R}}\left[1+\frac{\kappa\lambda}{\eta}\tau_{RC}\tau_{rr}^*\right]= \frac{\eta}{\lambda \tau_{ZC}}\frac{\partial \upsilon_r^*}{\partial z^*}  
\end{eqnarray}
From eq (3) and eq(4) it is observed that the $\tau_{RC}:=\eta/\lambda$, $\tau_{ZC}:=\eta/\lambda$. By substituting these scales into eq (2), the dimensionless radial momentum equation is deduced as represented in eq(5). 

\begin{eqnarray}
\frac{\rho u_c^2 \lambda}{\eta}\big( \frac{\partial \upsilon_{r}^*}{\partial t^*}+\upsilon_r^*\frac{\partial \upsilon_r^*}{\partial r^*} \big)=-\frac{\sigma\lambda}{\eta R_o}\frac{\partial p^*}{\partial r^*}+\frac{\tau_{rr}^*}{r^*}+\frac{\partial\tau_{rr}^*}{\partial r^*}+\frac{\partial\tau_{rz}^*}{\partial z^*}  
\end{eqnarray}

\begin{figure*}[h!]
\includegraphics[scale=0.39]{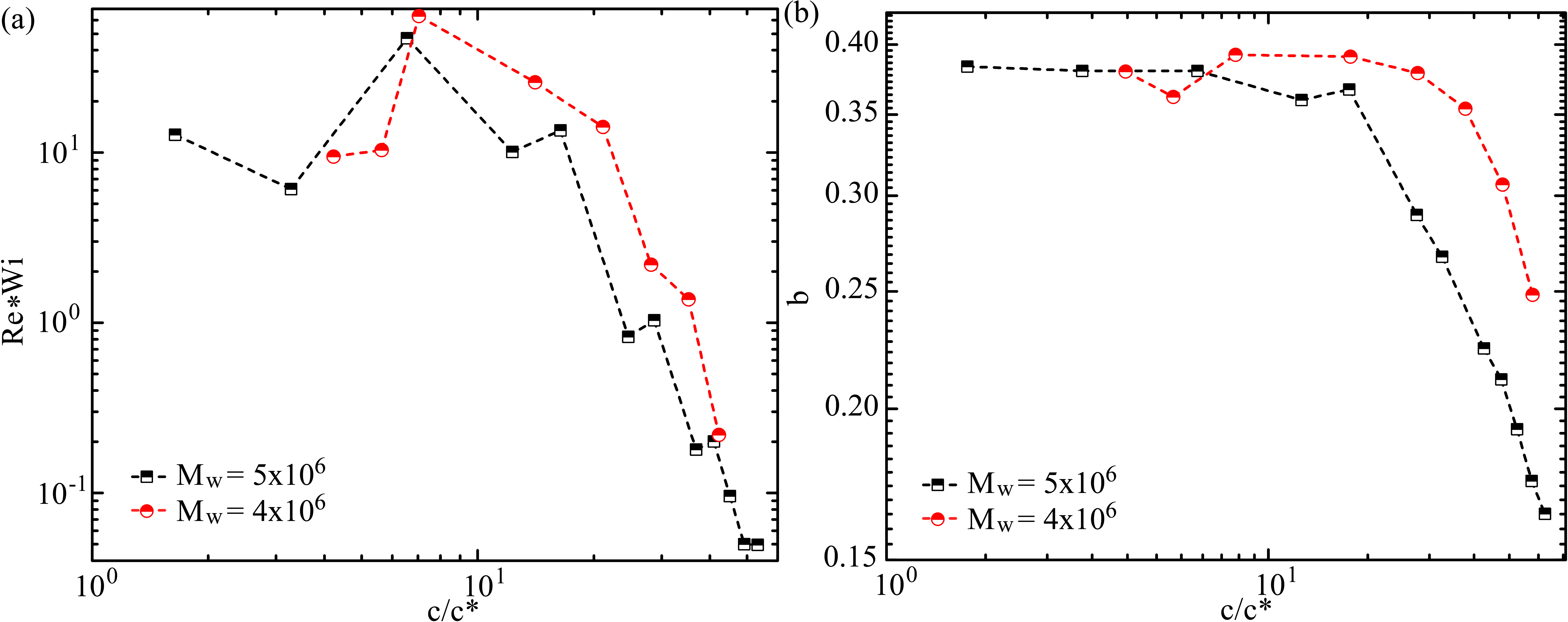}
\caption{(a) Variation of $Re*Wi$ with $c/c^*$ representing the decrease in $Re*Wi$ from $\mathcal{O}({10^1})$ to $<\mathcal{O}({10^0})$ from inertio-elastic/viscoelastic to elasticity dominated regime, (b) Representing the steady and monotonic decrease of $b$ in inertio-elastic/viscoelastic regimes ($c/c^*<20$) and elasticity dominated regime ($c/c^*>20$) respectively.}
\label{fig:numbers_b} 
\end{figure*}

The coefficient $\frac{\rho u_c^2 \lambda}{\eta}$, of inertial term in eq (5) is the product of $Re$ and $Wi$, which is given as $Re*Wi=<{\rho u_c^2 \lambda}/{\eta}>=\frac{\textrm{Elastic Force . Inertia Force}}{\textrm{{(Viscous Force)}}^2}$. The term $Re*Wi$ can be rewritten as $Re*Wi=<{{u_c^2}/{U_s^2}}>$ where $U_s=\sqrt{{\eta}/{\rho \lambda}}$\cite{75} is the shear wave velocity of the complex fluid. The values of $Re*Wi$ are presented for different concentration ratios in the Fig. 5(a). In the elasticity dominated regime, as observed in Fig. 4, the product of $Re$ and $Wi$ is $Re*Wi=<u_c^2/U_s^2><\mathcal{O}(10^0)$ while, for the other regimes $Re*Wi\sim\mathcal{O}(10^1)$. This implies that for the elasticity dominated regime, the characteristic  velocity of the system $u_c$ is less than the shear wave velocity of the fluid $U_s$ while $u_c>U_s$ for the other regimes. The polymer chains begin to elongate along the shear direction after the droplets have touched each other. Such elongation decreases as the concentration of polymer increases due to polymer chain entanglements which alter the curvature of the liquid bridge, leading to the slow growth of the bridge and inhibiting the coalescence. In the elasticity dominated regime, the chains relax slower than the speed of information transfer hence the polymer chains are in unrelaxed state. On the contrary, for the other regimes, the polymer chains relax faster than the speed of information transfer implying that the chains have already relaxed to the external perturbation. This behaviour of chain relaxation leading to the decrease of exponent $b$ in elasticity dominated regime where, $Re*Wi<\mathcal{O}(10^0)$ is represented in Fig. 5(b).    

\begin{figure*}[h!]
\includegraphics[scale=0.65]{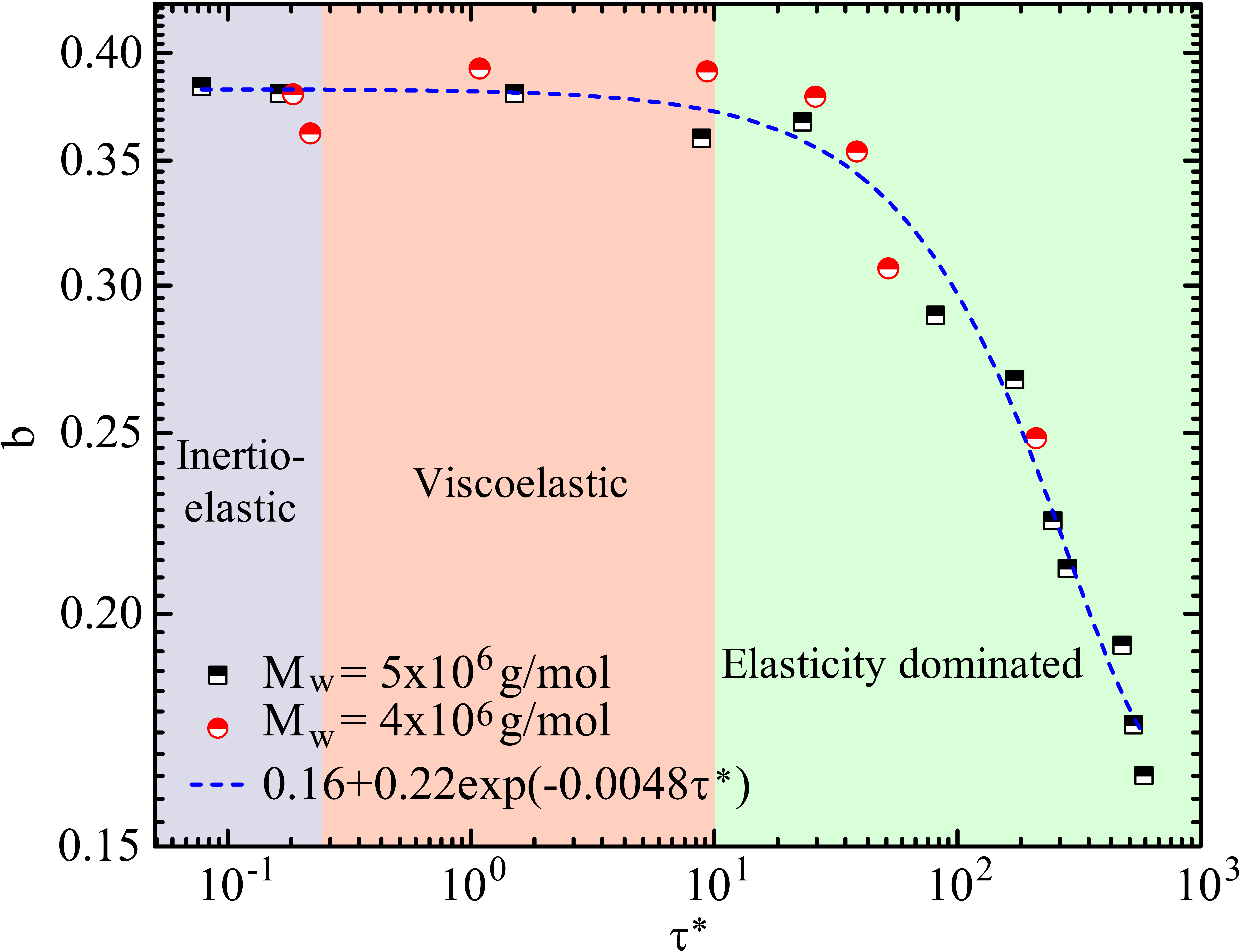}
\caption{Dependence of the power law index $b$ on the $\tau^*$ which is the ratio of relaxation time $\lambda$ and Newtonian characteristic time $t_c$, with dashed blue line representing the exponential fit of 97\% confidence interval for PEO solutions of different molecular weights.}
\label{fig:numbers_b} 
\end{figure*}

The effect of chain relaxation time is further demonstrated by considering the characteristic velocity $u_c$ as the chain relaxation velocity, which can be defined as $u_c=R_o/\lambda$. On substitution, the term ${\rho u_c^2 \lambda}/{\eta}$ can be simplified as ${\rho R_o^2}/{\eta\lambda}$. This simplified result can be rewritten as the ratio of time scales ($\tau^*$) ${\rho R_o^2}/{\eta\lambda}={t_c}/{\lambda}=1/\tau^{*2}$ where, $t_c=t_i^2/t_v$ is the Newtonian characteristic time. Here, $t_v={\eta R_o}/{\sigma}$ is the viscous time scale and $t_i=\sqrt{{\rho {R_o}^3}/{\sigma}}$ is the inertial time scale. It is observed from Fig.~\ref{fig:numbers_b} that when $\tau^*<10$, the exponent is constant with a value of 0.37 i.e the process is in inertio-elastic or viscoelastic regime. On contrary, when $\tau^*>10$, the exponent decreases continuously. The dynamics governing the above phenomenon lies in the relaxing of polymer chains. When $\tau^*<10$ the polymer chain relaxation are comparable to the Newtonian time scale $t_c$ leading to a constant value. However, for $\tau^*>10$, the polymer chains are in unrelaxed state even after the Newtonian time scale therefore altering the curvature of the bridge, resulting in the decline of $b$. 

The universal behaviour of the neck radius evolution is proposed by Sarath et al.\cite{our} in inertio-elatic regime. To attain the universality Sarath et al.\cite{our} non dimensionalized the neck radius $R$ using $\sqrt{\nu_o\lambda}$ as $\displaystyle R^*={R}/{\sqrt{\nu_o\lambda}}$. Similarly, time $t$ is non-dimensionalized with $\lambda{\textrm{Oh}^{-1}}(\frac{c}{c^*})^{-1.2}$ leading to $t^*=(\frac{t}{\lambda}{\textrm{Oh}^{-1}})(\frac{c}{c^*})^{-1.2}$. This non-dimensionalization led to the universal behaviour of the neck radius growth as $R^*\sim {t^*}^{0.36}$ which is in agreement for the solutions in inertio-elastic and viscoelastic regimes having a constant value of $b=0.37$ as represented in Fig. 6. However, such non-dimensionalization of neck growth breaks in the elasticity dominated regime. As the low $Wi$ assumption is no longer valid, the previously reported governing equations is unable to capture the deviation from universality. This deviation shown in the inset of Fig. 7 is due to the incorrect characteristic length and time obtained from the balance of inertia, elastic and capillary forces, as the inertial forces are weak in the elasticity dominated regime. Moreover, in this regime, the polymer chains are not relaxed, hence the temporal variation of stress in the upper convected derivative should be considered, which was neglected in previous studies\cite{our}. 

\begin{figure*}[h]
\includegraphics[scale=0.65]{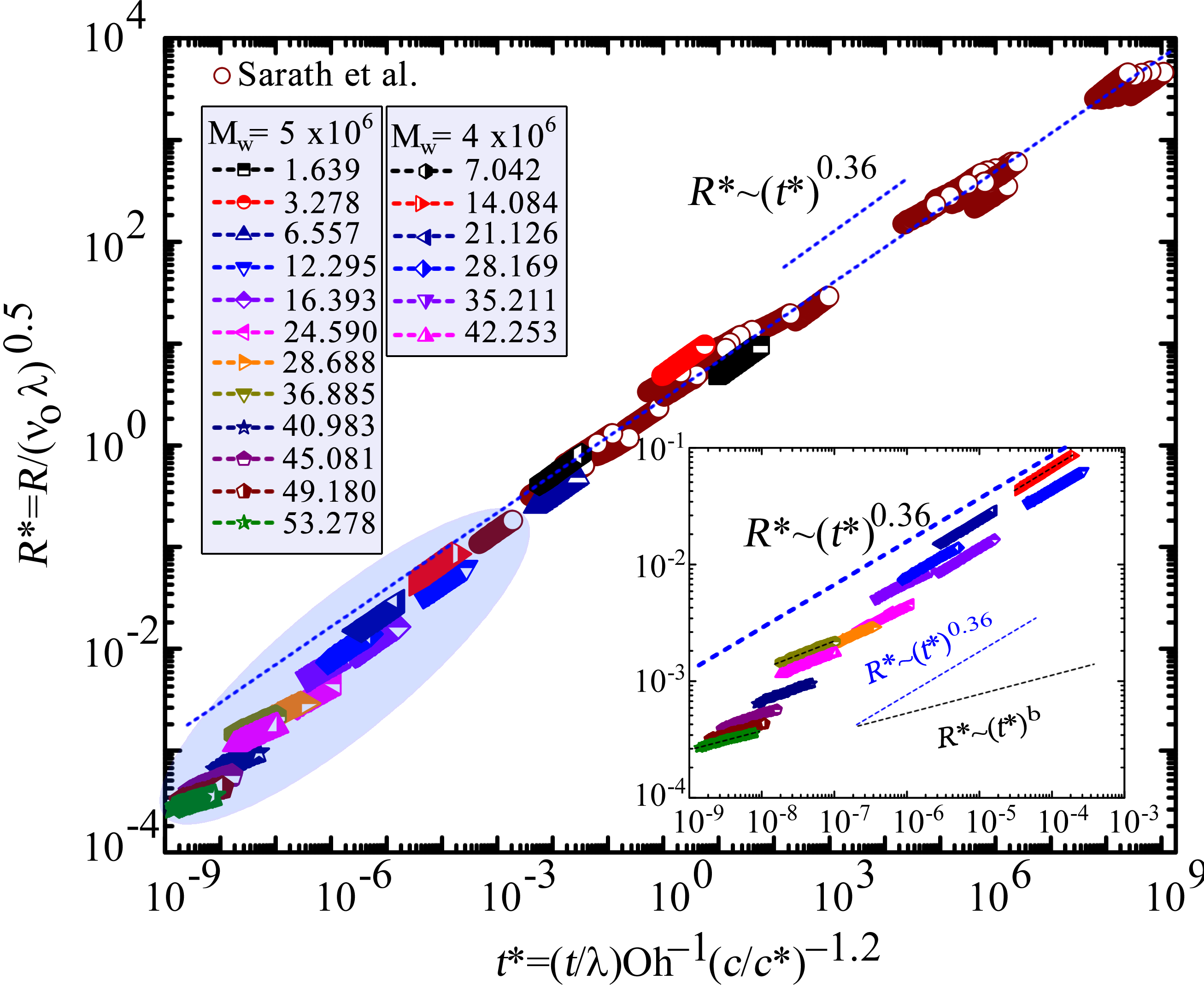} 
\caption{Non-dimensional neck radius as a function of non-dimensional time for all the polymer solutions used in this study and the previous study by Sarath.et.al.\cite{our} with legend representing corresponding $c/c^*$ values. Inset shows the breaking of universality for $c/c^*>10$ highlighted in the shaded region.}
\label{fig:universal} 
\end{figure*}

This deviation from universality in the elasticity dominated regime provides a novel method to determine the relaxation time $\lambda$ of the complex fluids using the coalescence experiment. From Fig. 6 we propose a correlation between $\tau^*$ and $b$ as $b=0.16+0.22\mathrm{exp}(-0.0048\tau^*)$ for PEO. This correlation for PEO is validated by conducting coalescence experiments for concentration ratio $c/c^*=32.56 (2.0\%  \mathrm{w/v})$ of PEO $M_w=5\times10^6$ solution having $\eta=55$ Pa.s. Under similar experimental conditions, the temporal evolution of the neck for $c/c^*=32.56$ is found to have the power-law exponent as $b=0.275$. On substituting $b=0.275$ in the correlation we obtain the relaxation time as $\lambda=0.6$ s, which agrees with the relaxation time obtained from the SAOS experiments $1.35$ s. Even though there is a difference in the relaxation times, it is known that the relaxation time is method specific. For instance, the relaxation time obtained from the Capillary breakup extensional rheometer (CABER)\cite{69,70,71,72} and SAOS differ by an order.

 In literature, many methods are proposed to measure the relaxation time of the fluid. Most widely used way to find $\lambda$ is the linear viscoelastic response in a conventional rheometer\cite{bird,68}. In this method, material is subjected to sinusoidal deformation to evaluate the viscous and elastic responses via loss modulus $G''$ and storage modulus $G'$ respectively. The crossover of $G'$ and $G''$ is used to determine the relaxation time of the fluid. But, this method is limited by the motor inertia in conventional rheometers and cannot capture the small values of $ \lambda$. Hence for low relaxation times, a novel method named CABER, Capillary breakup extensional rheometer- Dripping on substrate (CABER-DOS)\cite{macro1,macro2} was proposed. However, the intrinsic difficulties in this method lies in the controlling of elongational flow. Such difficulties have led to the significant difference in relaxation times measured from the conventional method and by CABER. Recently, there are developments in microfluidic devices\cite{74,66,73} for overcoming the limitations of conventional rheometer and CABER, but, the fabrication of the microfluidic channel is intricate. However, the present study proposes a simple comprehensive tool named Rheocoalescence based on empirical correlations to determine the relaxation time of PEO solutions. Even though the proposed correlation can be used in all the regimes, it is robust in elasticity dominated regime. A comprehensive study on this method is required to generalize Rheocoalescence for all the polymeric fluids. The required experimental information for the correlation can be obtained easily, which makes this tool predominantly effective for cases where performing experiments by conventional methods become very difficult such as, the case of highly elastic fluid. This technique opens up a new paradigm in microfluidics and rheological measurements.

\section{Conclusion}

The current study demonstrates the effect of fluid elasticity on coalescence of pendant-sessile polymeric droplets. We performed high speed imaging to capture the temporal evolution of the bridge for a wide range of concentrations ratios. We reveal the presence of three regimes namely inertio-elastic, viscoelastic and elasticity dominated regimes based on $c/c^*$. The inertio-elastic regime occurs at $c/c^*<c_e/c^*$, and viscoelastic regime at $c_e/c^*<c/c^*<c_c/c^*$, similarly elasticity dominated regime at $c/c^*>c_c/c^*$. Experimentally, we have been able to demonstrate the dependence of power law index $b$ on relaxation time leading to a novel method: Rheocoalescence to determine the relaxation time of the fluids. This opens a new paradigm in determining the characteristic time scales for wider class of complex fluids. However, the current study neglects the effect of surrounding fluid on the dynamics by considering air as the outer fluid. Further studies should be dedicated to extending this method's applicability for a variety of fluids along with the effect of outer fluid.

%%%%%%%%%%%%%%%%%%%%%%%%%%%%%%%%%%%%%%%%%%%%%%%%%%%%%%%%%%%%%%%%%%%%%
%% The "Acknowledgement" section can be given in all manuscript
%% classes.  This should be given within the "acknowledgement"
%% environment, which will make the correct section or running title.
%%%%%%%%%%%%%%%%%%%%%%%%%%%%%%%%%%%%%%%%%%%%%%%%%%%%%%%%%%%%%%%%%%%%%
\begin{acknowledgement}

\end{acknowledgement}

%%%%%%%%%%%%%%%%%%%%%%%%%%%%%%%%%%%%%%%%%%%%%%%%%%%%%%%%%%%%%%%%%%%%%
%% The same is true for Supporting Information, which should use the
%% suppinfo environment.
%%%%%%%%%%%%%%%%%%%%%%%%%%%%%%%%%%%%%%%%%%%%%%%%%%%%%%%%%%%%%%%%%%%%%
\begin{suppinfo}

\end{suppinfo}

%%%%%%%%%%%%%%%%%%%%%%%%%%%%%%%%%%%%%%%%%%%%%%%%%%%%%%%%%%%%%%%%%%%%%
%% The appropriate \bibliography command should be placed here.
%% Notice that the class file automatically sets \bibliographystyle
%% and also names the section correctly.
%%%%%%%%%%%%%%%%%%%%%%%%%%%%%%%%%%%%%%%%%%%%%%%%%%%%%%%%%%%%%%%%%%%%%
\bibliography{achemso-demo}

\end{document}